\documentclass[reprint,amsmath,amssymb,aps,prb]{revtex4-1}
\usepackage{amsmath,epsfig,wrapfig,graphicx,bm}

\begin{document}
\title{Magnetic State Selected by Magnetic Dipole Interaction in Kagome
Antiferromagnet NaBa$_{2}$Mn$_{3}$F$_{11}$}

\author{Shohei Hayashida,$^{1}$ Hajime Ishikawa,$^{2}$ Yoshihiko Okamoto,$^{3}$ 
Tsuyoshi Okubo,$^{4}$ Zenji Hiroi,$^{1}$ Maxim Avdeev,$^{5,6}$ Pascal Manuel,$^{7}$
Masato Hagihala,$^{1}$ Minoru Soda,$^{1}$ and Takatsugu Masuda$^{1}$}

\affiliation{$^{1}$Institute for Solid State Physics, The University of Tokyo, Chiba 277-8581, Japan\\
 $^{2}$Institute for Functional Matter and Quantum Technologies, University of Stuttgart, 
Pfaffenwaldring 57, 70569 Stuttgart, Germany\\
 $^{3}$Department of Applied Physics, Nagoya University, Nagoya 464-8603, Japan\\
$^{4}$Department of Physics, The University of Tokyo, Tokyo 113-0033, Japan \\
$^{5}$Australian Nuclear Science and Technology Organization, Menai, NSW 2234, Australia\\
$^{6}$School of Chemistry, The University of Sydney, Sydney, NSW 2006, Australia\\
$^{7}$ISIS Pulsed Neutron and Muon Source, Rutherford Appleton Laboratory, Chilton, Didcot OX11 0QX, United Kingdom}

\date{\today}

\begin{abstract}
We have studied the ground state of the classical Kagome antiferromagnet
NaBa$_{2}$Mn$_{3}$F$_{11}$. 
Strong magnetic Bragg peaks observed for $d$-spacings shorter than 6.0 \AA\ were 
indexed by the propagation vector of $\bm k_{0} = (0,0,0)$. 
Additional peaks with weak intensities in the $d$-spacing range above 8.0 \AA\  
were indexed by the incommensurate vector of 
$\bm k _{1}=(0.3209(2),0.3209(2),0)$ and $\bm k _{2}=(0.3338(4),0.3338(4),0)$. 
Magnetic structure analysis unveils a 120$^{\circ}$ structure with the 
{\it tail-chase} geometry having $\bm k_0$ modulated by the incommensurate vector. 
A classical calculation of the Kagome Heisenberg antiferromagnet 
with antiferromagnetic 2nd-neighbor interaction, for which the ground state 
a $\bm k_0$ 120$^{\circ}$ degenerated structure,
reveals that the magnetic dipole-dipole (MDD) interaction including up to the 4th neighbor 
terms selects the tail-chase structure. 
The observed modulation of the tail-chase structure is attributed to a small perturbation 
such as the long-range MDD interaction or the interlayer interaction. 
\end{abstract}

\maketitle
\section{Introduction}
%1st paragraph
Long-range magnetic dipole-dipole (MDD) interaction is ubiquitous in nature. 
The texture of iron fillings around a bar magnet is a visualization of the MDD 
interaction which is familiar to schoolchildren, and the anisotropic deformation of condensed
 magnetic atoms at a low temperature is at the forefront of modern science~\cite{Stuhler05}. 
In insulating magnets, 
effective quantum spins having large magnitude of moments coupled via the MDD 
give easy access to observations of novel quantum phenomena~\cite{Ronnow05,Kraemer12,Buruzuri11}. 
In artificial mesomagnets the vortex cores dominated by the long-range MDD 
exhibit complex collective dynamics in magnonic 
crystals~\cite{Sugimoto11,Jain12,Hanze16}.  
In bulk magnets composed of 3$d$ transition metals, however, the MDD 
is not necessarily a primary interaction but a small liaison to transfer the information 
of the lattice symmetry to the spin space. 
Luttinger and Tisza successfully explained several types of magnetic structures 
by the MDD in their pioneering work~\cite{PR70},
and several experimental studies followed~\cite{PR110_MnO,PRB86_MnO,PRB93_MnF2}.

%2nd paragraph
The MDD interaction is even more important in geometrically frustrated magnets,  
where the geometry causes macroscopic degeneracy.
% and the MDD induces unconventional magnetic states.
For instance, $A_{2}B_{2}{\rm O}_{7}$ pyrochlore oxides exhibiting MDD
display exotic states which are doubly gauge charged emergent magnetic
monopoles~\cite{RMP82}.
In an artificial magnet, collections of nanomagnetic islands arranged in a Kagome lattice 
generate magnetic moment fragmentation~\cite{RMP85}.
The combination of the frustrated geometry and the MDD interaction is thus 
a good playground for a new magnetic state.

\begin{figure}[htbp]
\epsfig{file=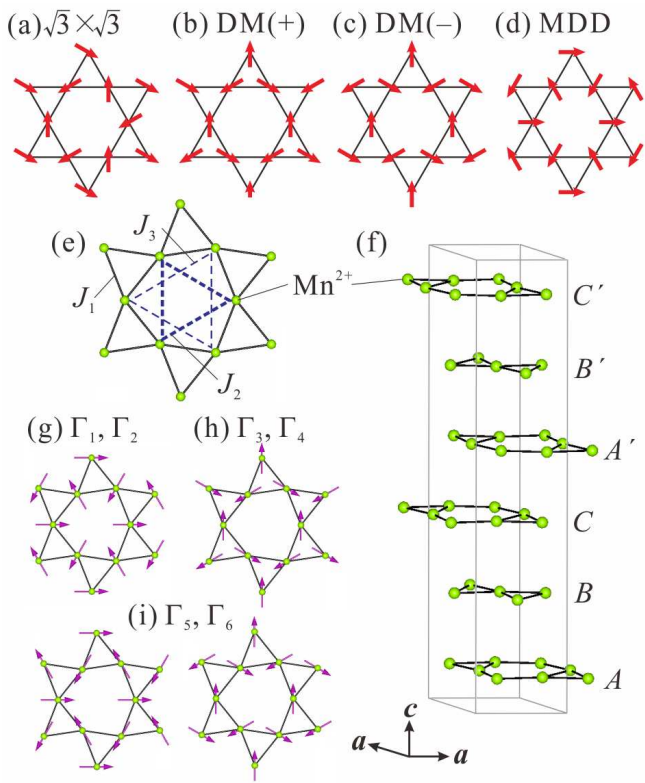,width=1.0\hsize}
\caption{120$^{\circ}$ structures in the Kagome lattice. The directions of the spins are 
represented by the red arrows. 
(a) 120$^{\circ}$ structure with the enlarged unit cell by $\sqrt{3}\times\sqrt{3}$.
(b) DM(+), (c) DM($-$), and (d) MDD-type
120$^{\circ}$ structure with the propagation vector $\bm k =0$.
(e) Mn$^{2+}$ ions in a Kagome layer in NaBa$_{2}$Mn$_{3}$F$_{11}$. 
Solid, thick-dashed and thin-dashed lines indicate the nearest-, second-, and third-neighbor
interactions.
The lattice is equivalent to the regular Kagome lattice as a spin system (see text).
(f) The linear perspective view of the Kagome layers.
(g)-(i) 120$^{\circ}$ structures represented by the IRs for $\bm k =0$.}
\label{fig1}
\end{figure}

%3rd paragraph
In a classical Heisenberg Kagome antiferromagnet, the ground state 
is infinitely degenerated. 
At zero temperature long-range order of the 120$^{\circ}$ structures with 
enlarged $\sqrt{3}\times\sqrt{3}$ unit cells characterized by ${\bm k}_{1/3} = (1/3, 1/3, 0)$ 
in Fig.~\ref{fig1}(a) is  
selected by the order-by-disorder  mechanism \cite{PRB48}.
The degeneracy of the ground state 
can be lifted also by various perturbations.  
The states selected by the Dzyaloshinskii-Moriya (DM) interaction 
are the 120$^{\circ}$ structures with $\bm k_0 = (0, 0, 0)$~\cite{PRB66};
the structures exhibit positive vector chirality in Fig.~\ref{fig1}(b) and negative vector 
chirality in Fig.~\ref{fig1}(c).
We name DM($+$)- and DM($-$)-type 120$^{\circ}$ structures for the former and the 
latter, respectively. 
The vector chirality is determined by the out-of-plane component of the DM vector. 
In the DM($+$) structure, the easy-axis anisotropy is induced by the in-plane component 
of the DM vector.  
The state selected by the MDD interaction is the 120$^{\circ}$ structure 
exhibiting {\it tail-chase} geometry as shown 
in Fig.~\ref{fig1}(d) \cite{PRB91}. 
It has positive chirality but different easy-axis anisotropy from the DM($+$) structure. 
It is named the MDD-type 120$^{\circ}$ structure. 
The structure is equivalent to magnetic vortices on a honeycomb lattice with staggered 
polarity, which can be a prototype of a natural magnonic crystal~\cite{Sugimoto11,Jain12,Hanze16}. 
The states selected by the second-neighbor interaction 
are the 120$^{\circ}$ structure with $\bm k_0$ for the antiferromagnetic case and 
that with $\bm k_{1/3}$ for the ferromagnetic case~\cite{PRB72_J2}.

%4th paragraph
The magnetic structures of the Kagome antiferromagnet 
have been intensively investigated by neutron diffraction 
on many compounds. 
The DM($+$) structure is realized in most 
cases; $A$Fe$_{3}$(SO$_{4}$)$_{2}$(OH)$_{6}$ ($A=$ K, Na, Ag, Rb, NH$_{4}$)~\cite{PRB61,PRB63,PRB61_DMp,PRB67_DMp}, 
KCr$_{3}$(SO$_{4}$)$_{2}$(OH)$_{6}$~\cite{PRB64}, and 
Nd$_{3}$Sb$_{3}$Mg$_{2}$O$_{14}$~\cite{PRB93}, 
which may be due to the coincidence between 
the direction of spins determined by DM interaction and the magnetic easy axis allowed by the crystallographic
symmetry. 
The DM($-$) structure is observed in a couple of semimetals Mn$_{3}$Sn and 
Mn$_{3}$Ge exhibiting large anomalous Hall effect~\cite{SSC42_DMm}. 
The $\sqrt{3} \times \sqrt{3}$ structure is found in the high pressure phase in 
herbertsmithite ZnCu$_{3}$(OH)$_{6}$Cl$_{2}$~\cite{PRL108_sqrt3}.
The tail-chase structure was observed in quinternary oxalate compounds with Fe$^{2+}$ 
ion~\cite{PRL107_MDD,EPJB_MDD} so far.
Its tail-chase structure was, however, caused by a strong single-ion anisotropy instead of
the MDD interaction.
To the best of our knowledge, the experimental observation of the tail-chase structure 
originating from the MDD interaction has not yet been identified (by neutron diffraction)
although it is of primary importance to the understanding of the Kagome family of compounds.

%5th paragraph
NaBa$_{2}$Mn$_{3}$F$_{11}$ crystallizes in a hexagonal structure with the space group
 $R\bar{3}c$~\cite{JSSC98}.
Mn$^{2+}$ ions carry spin $S=5/2$, and MnF$_{7}$ pentagonal bipyramids form a 
Kagome lattice in the crystallographic $ab$-plane as shown in Fig.~\ref{fig1}(e).
The path of the nearest-neighbor interaction $J_{1}$ indicated by the solid line is 
Mn-F-Mn. 
Although the interior angles of the hexagon in the Kagome lattice are shifted from 
120$^{\circ}$ and the lattice is distorted,
the length of the sides and the angles of Mn-F-Mn are the same for all the bonds.
This means that the magnitudes of the nearest-neighbor interactions are the same.
The spin system is thus regarded as the regular Kagome lattice. 
The six Kagome layers are stacked in the unit cell as shown in Fig.~\ref{fig1}(f). 
The $A$, $B$, and $C$ layers and $A^{\prime}$, $B^{\prime}$, and $C^{\prime}$
 layers are related by the $c$-glide.

The exchange pathways of the second and third-neighbor interaction are unusual;
the second-neighbor interaction $J_{2}$ indicated by the thick dashed line is Mn-F-Mn, 
and that of the third-neighbor interaction $J_{3}$ 
indicated by the thin dashed line is Mn-F-F-Mn.
The $J_{3}$ is thus negligible, and the unique network called Kagome-Triangular lattice
is realized \cite{JPSJ83}.
The heat capacity and magnetic susceptibility
measurements exhibit antiferromagnetic transition at $T_{{\rm N}}=2$ K.
The Curie-Weiss temperature $\theta_{{\rm CW}}$ was estimated
to be $-32$ K, which is smaller than those of most Kagome lattice magnets~\cite{PRB61,PRB63,PRB61_DMp,PRB67_DMp,PRB64,PRB93,PRL108_sqrt3}. 
In addition, the bond angles of the nearest neighbor exchange pathways are close to
90$^{\circ}$ rather than 180$^{\circ}$ \cite{JPSJ83},
suggesting the nearest neighbor interaction is weak antiferromagnet or ferromagnetic
based on the Goodenough-Kanamori rules \cite{goodenough,kanamori}.
The exchange interaction in NaBa$_{2}$Mn$_{3}$F$_{11}$ is thus 
relatively small, and the MDD interaction may be important.   

In this paper, we demonstrate that 
the tail-chase structure with small incommensurate (IC) modulations is realized
 in NaBa$_{2}$Mn$_{3}$F$_{11}$ by using neutron diffraction. 
Combination of the experiment and calculation suggests that 
the tail-chase structure selected by the main perturbation of the short-range MDD interaction including up to the fourth neighboring 
is modulated by a smaller perturbation 
such as the long-range MDD interaction or the interlayer interaction.

\section{Experimental details}
%6th paragraph
A polycrystalline sample was prepared by a solid state reaction method~\cite{JPSJ83}. 
The total mass of the obtained sample was 5.4 g.
A $^{3}$He cryostat was used to achieve low temperatures.
Neutron diffraction experiments were performed using two neutron diffractometers; 
a powder diffractometer ECHIDNA installed in OPAL reactor, Australian Nuclear Science
and Technology Organization for the preliminary measurement, and 
a long-wavelength time-of-flight (TOF) diffractometer WISH~\cite{wish} installed at the
ISIS Pulsed Neutron and Muon Source, Rutherford Appleton Laboratory 
for the precise measurement.
We chose a high resolution double frame mode at WISH.
The data for the Rietveld refinement in Figs.~\ref{fig2}(a), \ref{fig2}(b) 
and the temperature dependence of the integrated intensities in Fig.~\ref{fig3}(a)  
were measured by using the detector bank with an average scattering angle of $2\theta=90^{\circ}$. 
The data for the diffuse scattering in Fig.~\ref{fig3}(b) were measured by using  
the detector bank centered at $2\theta=27^{\circ}$.
The obtained data were analyzed by the Rietveld method using FullProf software~\cite{PhysicaB192}.  
Candidates for the magnetic structure compatible with the lattice symmetry were obtained
by the SARA$h$ software \cite{PhysicaB276}.

\section{Results and analysis}
%7th paragraph
\begin{figure}[tbp]
\epsfig{file=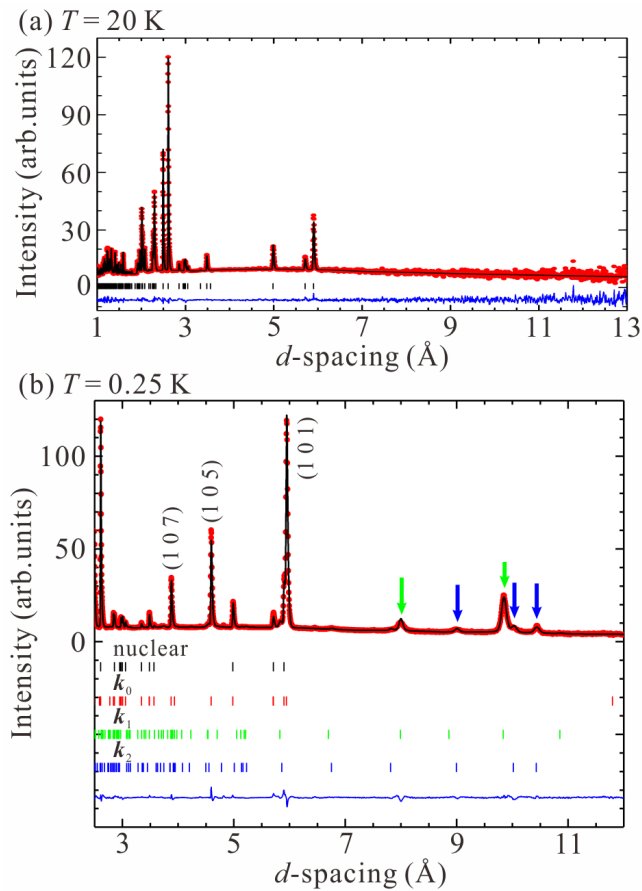,width=1.0\hsize}
\caption{Neutron diffraction profiles for NaBa$_{2}$Mn$_{3}$F$_{11}$ at (a) 20 K
and (b) 0.25 K.
The solid squares and curves show the experimental data and simulations, respectively. 
The vertical bars show the positions of the nuclear and 
magnetic Bragg peaks.
The solid curves below the bars show the difference between the data and simulations.
The green and blue arrows indicate the IC magnetic Bragg 
peaks with $\bm k _{1}$ and $\bm k _{2}$.}
\label{fig2}
\end{figure}

The neutron diffraction profile measured at 20 K is reasonably fitted by the hexagonal
structure with the space group $R\overline{3}c$ as shown in Fig.~\ref{fig2}(a). 
The profile factors are $R_{{\rm wp}}=8.80\%$ and $R_{{\rm e}}=4.37\%$,  
and the obtained parameters are summarized in the cif file in supplemental information. 

\begin{figure}[tbp]
\epsfig{file=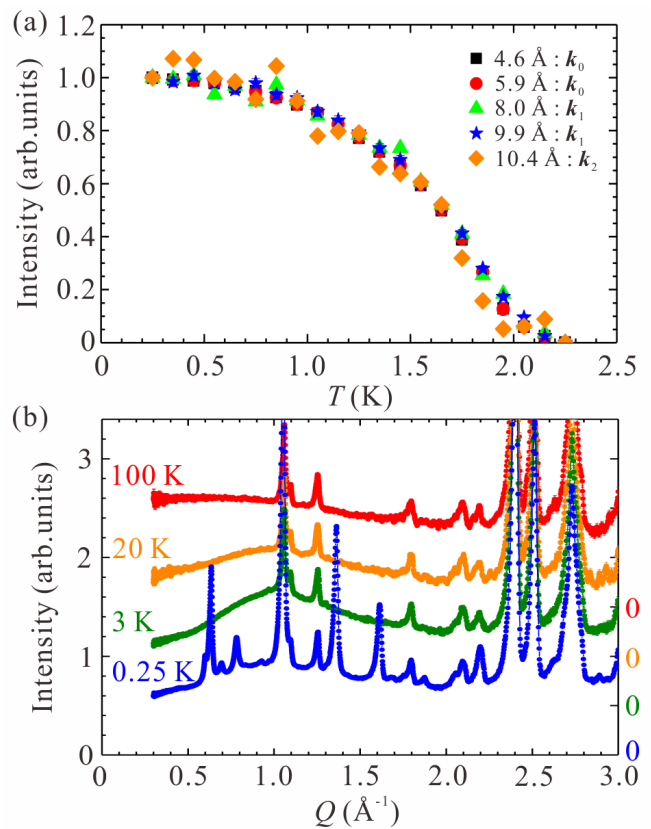,width=1.0\hsize}
\caption{(a) Temperature evolution of the integrated intensities at 
$d=$ 4.6, 5.9, 8.0, 9.9 and 10.4 {\AA}. 
The error bars are inside the markers.
Each of integrated intensities is normalized to their values at 0.25 K and subtracted 
by their background at 2.25 K.
(b) Neutron diffraction profiles at $T=$ 100, 20, 3 and 0.25 K.
The profiles are shifted by vertical offsets.}
\label{fig3}
\end{figure}

%8th paragraph
At 0.25 K, at least eight additional peaks are observed as shown in Fig.~\ref{fig2}(b).
The peak intensities increase with the decrease of the temperature below 2.25 K as 
shown in Fig.~\ref{fig3}(a).
This means that a magnetic long range order occurs at $T_{\rm N}=2.25$ K,
which is consistent with the previous heat capacity measurement~\cite{JPSJ83}.
The peaks at $d=$ 3.8, 4.6, and 5.9 {\AA} are indexed as (1~0~7),  (1~0~5), 
and (1~0~1), meaning that the magnetic propagation vector is $\bm k _{0}=(0,0,0)$.
The peaks indicated by the green and blue arrows in the long $d$ region are not indexed
by the $\bm k _{0}$ but by IC vectors.

%9th paragraph
Temperature variation of the diffraction profiles are exhibited in 
Fig.~\ref{fig3}(b). 
At 100 K paramagnetic scattering is observed in the low $Q$ region. 
On cooling it is suppressed, and, instead, magnetic diffuse scattering is 
induced at $Q \sim $ 1.0 {\AA}$^{-1}$ and more pronounced at 3 K. 
The diffuse scattering is suppressed with 
further cooling, and magnetic Bragg peaks appear. 
The short-range spin correlations thus develop at much higher temperature 
than the transition temperature, suggesting the existence of strong geometrical frustration. 
The behavior is consistent with the heat capacity in which most of the magnetic entropy was
released above $T_{{\rm N}}$~\cite{JPSJ83}.

%10th paragraph
In the magnetic structure analysis, it is assumed that the peaks with $\bm k _{0}$ mainly 
construct the magnetic structure, since the intensities of the peaks with $\bm k _{0}$ are
larger than those with the IC vectors.
The representation analysis~\cite{PhysicaB276} with the space group $R\bar{3}c$ 
and the propagation vector $\bm k _{0}$ leads to six irreducible representations (IRs)
 $\Gamma_{1}+\Gamma_{2}+\Gamma_{3}+\Gamma_{4}+2\Gamma_{5}+2\Gamma_{6}$.
The IRs and the basis vectors are summarized in Table \ref{tb:IRq0}.  
The basis vectors for $\Gamma_{1}$ or $\Gamma_{2}$ provide the MDD-type 
120$^{\circ}$ structure in Fig.~\ref{fig1}(g),  
and $\Gamma_{3}$ or $\Gamma_{4}$ provide the DM($+$)-type structure in Fig.~\ref{fig1}(h) whereas the basis vectors associated with $\Gamma_{5}$ or $\Gamma_{6}$
correspond to the 120$^{\circ}$ structure with the negative vector chirality as shown in Fig.~\ref{fig1}(i).
The magnetic structure in the ${\alpha}$ layer (${\alpha} = A, B, C$) and 
that in the ${\alpha}'$ layer (${\alpha}' = A', B', C'$) are the same for  $\Gamma_{1}$, $\Gamma_{3}$ and $\Gamma_{5}$. 
In contrast, the structure in the ${\alpha}'$ layer 
is the inversed structure of the ${\alpha}$ layer 
for $\Gamma_{2}$, $\Gamma_{4}$, and $\Gamma_{6}$.  
In testing the models of the magnetic structures inferred by the various IRs,
it is assumed that the magnitude of the magnetic moments on the Mn$^{2+}$ ions 
are all the same.
From the Rietveld refinements, we find that 
only $\Gamma_{2}$ gives a satisfactory agreement with the observed pattern.
The refined magnetic structure with $\bm k _{0}$ exhibits 
the 120$^{\circ}$ structure in the $ab$-plane
as shown in Fig.~\ref{fig1}(g). 
The refined magnitude of the moment is 
4.14(1) $\mu_{{\rm B}}$ at 0.25 K, which is 83$\%$ of the full moment of Mn$^{2+}$ ion.
According to the $J_{1}$ - $J_{2}$ phase diagram in the Heisenberg 
Kagome-Triangular lattice~\cite{JPSJ83}, 
the 120$^{\circ}$ structure with $\bm k _{0}$ is favored in case that both of $J_{1}$
 and $J_{2}$ are antiferromagnetic. 
This means that both of $J_{1}$ and $J_{2}$ in this compound are
antiferromagnetic in the absence of MDD interaction.

%11th paragraph
We search the propagation vectors of the IC peaks corresponding high symmetry
points/lines/planes of the Brillouin Zone.
The IC peaks are indexed by two propagation vectors: 
$\bm k _{1}=(0.3209(2),0.3209(2),0)$ for the peaks at $d=$ 8.0 and 9.9~{\AA} 
and $\bm k _{2}=(0.3338(4),0.3338(4),0)$ for those at $d=$ 9.0, 10.0 and 10.4~{\AA}. 
The IC vectors are close to ${\bm k}_{1/3} = (1/3, 1/3, 0).$
The representation analysis with the space group $R\bar{3}c$ and the propagation vectors
 $\bm k _{1}$ and $\bm k _{2}$ leads to separation of the equivalent Mn sites into the four
nonequivalent Mn sites, and two IRs $\Gamma_{1}+\Gamma_{2}$
at each of the four Mn sites. 
The IRs and the basis vectors are summarized in Table \ref{tb:IRIC}.  
We construct the models of the magnetic structure by the linear combinations of 
the basis vectors in each single IR. 
The explicit formulas of the magnetic models that are compatible with both propagation
vectors and the space group in the case of $\Gamma_{2}$ are as follow:
\begin{eqnarray}
\bm m _{{\rm Mn1a}}&=&c_{4}^{(1)}\Psi_{4}^{(1)}+c_{5}^{(1)}\Psi_{5}^{(1)}+c_{6}^{(1)}\Psi_{6}^{(1)}, \\
\bm m _{{\rm Mn1b}}&=&c_{4}^{(1)}\Psi_{4}^{(1)}+c_{5}^{(1)}\Psi_{5}^{(1)}+c_{6}^{(1)}\Psi_{6}^{(1)}, \\
\bm m _{{\rm Mn2}}&=&c_{2}^{(2)}\Psi_{2}^{(2)}+c_{3}^{(2)}\Psi_{3}^{(2)}, \\
\bm m _{{\rm Mn3a}}&=&(c_{4^{\prime}}^{(1)}\Psi_{4}^{(3)}+c_{5^{\prime}}^{(1)}\Psi_{5}^{(3)})
+c_{6^{\prime}}^{(1)}\Psi_{6}^{(3)}, \\
\bm m _{{\rm Mn3b}}&=&(c_{4^{\prime}}^{(1)}\Psi_{4}^{(3)}+c_{5^{\prime}}^{(1)}\Psi_{5}^{(3)})
+c_{6^{\prime}}^{(1)}\Psi_{6}^{(3)}, \\
\bm m _{{\rm Mn4}}&=&c_{2^{\prime}}^{(2)}\Psi_{2}^{(4)}+c_{3^{\prime}}^{(2)}\Psi_{3}^{(4)}. 
\end{eqnarray}
Here the coordinations of the Mn atoms and the basis vectors $\Psi_{i}^{(j)}$ are 
exhibited in Table~\ref{tb:IRIC}.
$c_{4}^{(1)}$, $c_{5}^{(1)}$, $c_{6}^{(1)}$, $c_{4^{\prime}}^{(1)}$, 
$c_{5^{\prime}}^{(1)}$, $c_{6^{\prime}}^{(1)}$, $c_{2}^{(2)}$, $c_{3}^{(2)}$,
$c_{2^{\prime}}^{(2)}$ and $c_{3^{\prime}}^{(2)}$ are 
coefficients of the linear combination of the basis vectors. 
The number of the fitting parameters is 10, which is too many for the number of the
observed IC peaks.
We therefore assumed that the magnetic structures in the six layers are as similar as 
possible, and the formulas used for the refinement were reduced to:
\begin{eqnarray}
\bm m _{{\rm Mn1a}}&=&c_{4}^{(1)}\Psi_{4}^{(1)}+c_{5}^{(1)}\Psi_{5}^{(1)}+c_{6}^{(1)}\Psi_{6}^{(1)}, \\
\bm m _{{\rm Mn1b}}&=&c_{4}^{(1)}\Psi_{4}^{(1)}+c_{5}^{(1)}\Psi_{5}^{(1)}+c_{6}^{(1)}\Psi_{6}^{(1)}, \\
\bm m _{{\rm Mn2}}&=&c_{2}^{(2)}\Psi_{2}^{(2)}+c_{3}^{(2)}\Psi_{3}^{(2)}, \\
\bm m _{{\rm Mn3a}}&=&t_{1}(c_{4}^{(1)}\Psi_{4}^{(3)}+c_{5}^{(1)}\Psi_{5}^{(3)})
+t_{2}c_{6}^{(1)}\Psi_{6}^{(3)}, \\
\bm m _{{\rm Mn3b}}&=&t_{1}(c_{4}^{(1)}\Psi_{4}^{(3)}+c_{5}^{(1)}\Psi_{5}^{(3)})
+t_{2}c_{6}^{(1)}\Psi_{6}^{(3)}, \\
\bm m _{{\rm Mn4}}&=&t_{1}c_{2}^{(2)}\Psi_{2}^{(4)}+t_{2}c_{3}^{(2)}\Psi_{3}^{(4)}. 
\end{eqnarray}
$t_{1}$ and $t_{2}$ take $+1$ or $-1$.
This reduces the number of parameters down to 5 and renders the refinement possible.
We similarly construct the magnetic models in the case of $\Gamma_{1}$.
The best fit is obtained for the $\Gamma_{2}$, and the parameters are listed in
Table~\ref{tb:refineIC}.
The profiles at 0.25 K and fitting results of the model combining 
with $\bm k _{0}$, $\bm k _{1}$, and $\bm k _{2}$ are shown in Fig.~\ref{fig2}(b).
The $R$ factors for the whole profile are $R_{{\rm wp}}=7.61\%$ and $R_{{\rm e}}=1.02\%$.
The magnetic $R$ factors $R_{{\rm mag}}$ for $\bm k _{0}$, $\bm k _{1}$,
and $\bm k _{2}$ are $2.71\%$, $7.62\%$, and $12.8\%$.
For the reference, the best $R_{{\rm mag}}$s for $\bm k _{0}$ are $57\%$ for 
$\Gamma_{1}$, $89\%$ for $\Gamma_{3}$, $80\%$ for $\Gamma_{4}$, 
$79\%$ for $\Gamma_{5}$, and $36\%$ for $\Gamma_{6}$.
The refined magnetic moments in each IC structure with $\bm k _{1}$ 
and $\bm k _{2}$ form the two-in-one-out (one-in-two-out) structure 
similar to the Kagome spin ice~\cite{PRB66_wills}. 
In addition, they have an out of plane component, the directions of which are all up or all
down, and the magnitudes of the moments are modulated.

%13th paragraph
The temperature evolutions of the integrated intensities associated with the propagation 
vectors $\bm k _{0}$, $\bm k _{1}$, and $\bm k _{2}$ are the same as shown in Fig.~\ref{fig3}(a).
This suggests that the low temperature state is a single ordered state, i.e., a 
multiple-$\bm k$ state, where the Mn$^{2+}$ moments form a 120$^{\circ}$ structure 
in the $ab$-plane, and the IC propagation vectors modulate this 120$^{\circ}$ structure.
The averaged magnitude of the magnetic moment of the Mn$^{2+}$ ion is 
4.54 $\mu_{{\rm B}}$ at 0.25 K, which is 91$\%$ of the Mn$^{2+}$ moment 
$(S=5/2)$.

\begin{figure}[tbp]
\epsfig{file=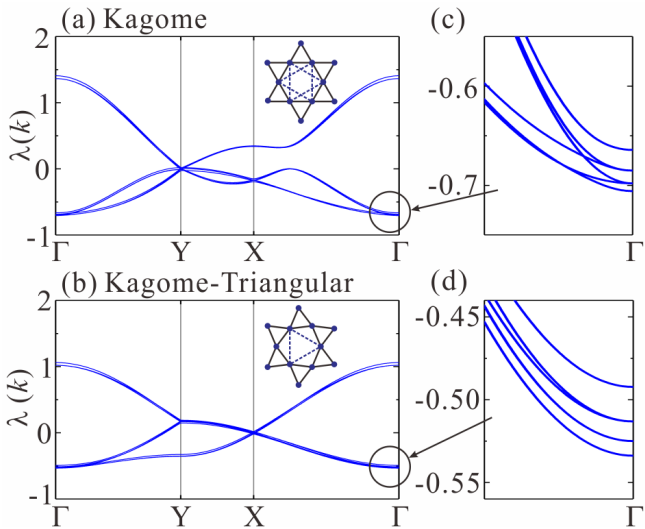,width=1.0\hsize}
\caption{Eigenvalues of the interaction matrix $J_{ij}^{\alpha\beta}$ along lines in the
Brillouin zone.
Spectra in (a) the Kagome lattice and (b) the Kagome-Triangular lattice where
the exchange interactions are $J_{1}=J_{2}=2$ K and the MDD interaction is $J_{{\rm MDD}}=56$ mK.
(c), (d) Detailed structures of the spectra around the $\Gamma$ point.}
\label{fig4}
\end{figure}

\section{Discussion}
%13th paragraph
For the calculation of the ground state we assume isotropic Heisenberg interactions, since the orbital angular momentum 
of Mn$^{2+}$ ion is zero, at least for the ground state of the isolated Mn$^{2+}$ ion, 
and the anisotropy and/or asymmetric terms derived from the perturbation of spin-orbit coupling 
should be small. 
As described in the introduction section, the geometry of the main framework of 
NaBa$_{2}$Mn$_{3}$F$_{11}$ is a Kagome-Triangular lattice and MDD interaction is not 
negligible. 
The following Hamiltonian in a Kagome plane is thus a good approximation for this system:
\begin{eqnarray}
&{\mathcal H}&=\sum_{{\rm n.n.}}J_{1}\bm S _{i}\cdot\bm S _{j}
+\sum_{{\rm n.n.n.}}J_{2}\bm S _{i}\cdot\bm S _{j} \nonumber \\
&+&\sum_{i,j}\frac{\mu_{0}}{4\pi}\frac{(g\mu_{{\rm B}})^{2}}{|\bm r _{ij}|^{3}}\left[\bm S _{i}\cdot\bm S _{j}
-3\frac{\left(\bm S _{i}\cdot\bm r _{ij}\right)\left(\bm S _{j}\cdot\bm r _{ij}\right)}
{|\bm r _{ij}|^{2}}\right], \nonumber \\
\label{hamiltonian}
\end{eqnarray}
where $J_{1}$ and $J_{2}$ are the exchange interactions in the nearest- and
second-neighbor paths. 
The third term is the MDD interaction up to the fourth-neighbor paths and 
$\bm r _{ij}$ is the bond vector between the spins.
The strength of the nearest neighbor MDD interaction $J_{{\rm MDD}}$ is 56 mK, 
which is determined from the distance of the nearest neighbor path $r_{{\rm n.n.}}$as follow:
\begin{equation}
J_{{\rm MDD}} \equiv \frac{\mu_{0}}{4\pi}\frac{(g\mu_{{\rm B}})^{2}}{r_{{\rm n.n.}}^{3}}=
56~{\rm mK}.
\end{equation}
In the calculation, the interlayer interaction is not included.
To calculate the ground state of the system, we use 
a Luttinger-Tisza-type theory~\cite{PR70}, and investigate the eigenenergies and 
eigenvectors of the interaction matrix in the wave vector space.
The Fourier transformed Hamiltonian can be written as
\begin{equation}
{\mathcal H}=\sum_{\bm k}\sum_{i,j}\sum_{\alpha,\beta}
J_{ij}^{\alpha\beta}(\bm k)S_{i}^{\alpha}(-\bm k)S_{j}^{\beta}(\bm k), 
\end{equation}
where $J_{ij}^{\alpha\beta}$ is the sum of  $J_{1}$, $J_{2}$, and $J_{{\rm MDD}}$.
Here, $\alpha$ and $\beta$ are the Cartesian indices of the spins and $i$, $j$ run
over the three basis sites in the unit cell of the Kagome lattice.
The spin vector is the Fourier component of the real space, and $\bm k$ runs over 
the Brillouin zone of the Kagome lattice.
Thus, for a given value of $\bm k$, $J_{ij}^{\alpha\beta}$ is a $9\times 9$ matrix that 
needs to be diagonalized. 
We calculate for two cases: Kagome lattice with second-neighbor interaction in 
Fig.~\ref{fig4}(a) and Kagome-Triangular lattice in Fig.~\ref{fig4}(b). 
$(0,0,0)$, $(1/3,1/3,0)$, and $(1/2,0,0)$ points of the Brillouin zone are labeled as
the $\Gamma$, X, and Y, respectively.
In order to realize the 120$^{\circ}$ structure with $\bm k=0$, we set 
antiferromagnetic interactions for both $J_{1}$ and $J_{2}$~\cite{PRB72_J2,JPSJ83}.
Since varying $J_{2}/J_{1}$ does not make a significant difference to the results
within a wide range of values, the exchange interactions are parametrized at $J_{1}=J_{2}$ for simplicity. 
We also put $J_{1}>J_{{\rm MDD}}$ because the Curie-Weiss temperature 
$\theta_{{\rm CW}}=-32$ K~\cite{JPSJ83} is larger than the $J_{{\rm MDD}}=56$ mK. 

The eigenenergy $\lambda(\bm k)$ is minimized for $\bm k =0$ in both lattices
 as shown in Figs.~\ref{fig4}(a) and \ref{fig4}(b) which implies that the MDD-type 120$^{\circ}$ structures in Fig.~\ref{fig1}(d) is realized for 
both Kagome and Kagome-Triangular lattices.
This result is consistent with the previous study~\cite{PRB91}.
Although the six states are degenerated at $\bm k =0$ in the absence of MDD interaction, 
the degeneracy is lifted by the interaction as shown in Figs.~\ref{fig4}(c) 
and \ref{fig4}(d).
The calculated ground states correspond to the magnetic structure having
 $\bm k _{0}$ in the experiment, but they do not reproduce the multiple-$\bm k$ structure. 

\begin{figure}[tbp]
\epsfig{file=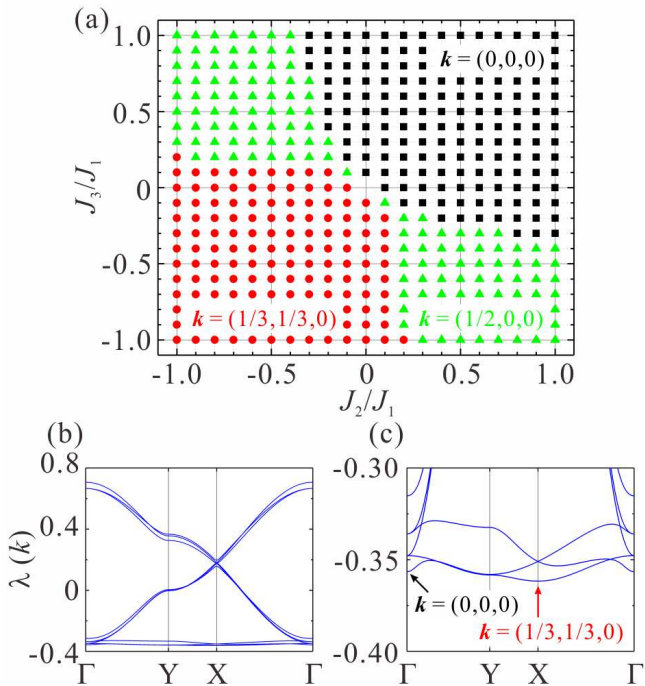,width=1.0\hsize}
\caption{(a) Phase diagram of Kagome-Triangular antiferromagnet 
having $J_3$ interaction. MDD interaction is not included. 
(b), (c) Eigenvalues of the interaction matrix $J_{ij}^{\alpha\beta}$ for 
Kagome-Triangular antiferromagnet having $J_3$ and MDD interaction along lines in the
Brillouin zone, where the parameters are $J_1$ = 2 K and  
$J_2 = J_3 = -J_{\rm MDD}/2$. The panel (b) is for wide energy range and the 
panel (c) is for low energy range. Double minima structure is observed. }
\label{fig5}
\end{figure}

For the multiple-$\bm k$ structure, we calculated the ground state of 
the Kagome-Triangular antiferromagnet including $J_3$ interaction in Fig.~\ref{fig1}(e). 
The obtained phase diagram of $J_3/J_1$ - $J_2/J_1$, where $J_1$ is antiferromagnetic, 
is shown in Fig.~\ref{fig5}(a). 
We have presumed that $J_1 \sim J_2 \gg |J_3|$ so far, and the observed 
$\bm k _{0}$ structure is confirmed by the calculation in this region. 
In case that $J_2$ and $J_3$ are ferromagnetic, i.e., in the third quadrant, 
the state of $\bm k_{1/3} = (1/3, 1/3, 0)$ which is close to the 
experimentally observed IC vectors $\bm k _{1}$ and $\bm k _{2}$ appears. 
The ground energy has, however, a single minimum in the $\bm k$ space, and the observed a multiple-$\bm k$ structure 
cannot be explained. 
Then the MDD interaction up to the 4th-neighbor paths is included, and the 
eigenvalues of the interaction matrix 
for $J_1$ = 2 K, $J_2$ = $J_3$ = $-J_{\rm MDD}/2$ 
is obtained as shown in Figs.~\ref{fig5}(b) and \ref{fig5}(c). 
Local minima appear at $\bm k = (1/3, 1/3, 0)$ and $(0, 0, 0)$, 
indicating the multiple-$\bm k$ structure. 
We found that the MDD interaction mixes the $\bm k _{0}$ structure in the first quadrant 
and $\bm k_{1/3}$ structure in the third quadrant in the  $J_3/J_1$ - $J_2/J_1$ phase diagram. 
The spin structure for $\bm k_0$ is the tail-chase structure which is consistent with the experiment. 
The one for $\bm k_{1/3}$ has solely out-of-plane component, and it exhibits up-up-down structure. 
This is inconsistent with the experimentally obtained structure. 
We have surveyed a series of parameters and exact match to the experimental structure 
could not be found.  
Thus, the terms in Eq.~(\ref{hamiltonian}) do not explain the observed multiple-$\bm k$ structure, and neither does $J_3$. 
Detailed theoretical studies considering further interactions including 
the long-range MDD interaction and/or the interlayer interaction
are necessary to reproduce the observed multiple-$\bm k$ 
structure including IC modulation for the future work.

%15th paragraph
The reason why the MDD interaction is the main perturbation 
in NaBa$_{2}$Mn$_{3}$F$_{11}$ is due to the fact that the exchange interaction is weak 
compared with most of Kagome antiferromagnets having O$^{2-}$ as anion that transfers 
the exchange integral \cite{PRB61,PRB63,PRB61_DMp,PRB67_DMp,PRB64,PRB93}.
Hybridization of the $d$ and $p$ orbitals is small in fluorides 
compared with oxides since covalency of F$^{-}$ ion is weaker than that of O$^{2-}$
ion.
In addition, the edge-sharing of the pentagonal bipyramids MnF$_{7}$ in the 
nearest neighbor path weakens the antiferromagnetic exchange interaction.
The superexchange interaction is thus weak 
and, consequently, DM interaction, that is resulting term of the perturbative treatment
of the exchange interaction and spin-orbit interaction in the Heisenberg spin Hamiltonian, is 
also weak. 
Furthermore, the charge distribution of Mn$^{2+}$ ion 
is spherical and prevents the appearance of single-ion anisotropy, 
since the $3d$ orbitals are half filled, with five electrons 
coupled giving rise to an angular momentum $L=0$. 
The MDD interaction hence causes the main perturbation  in NaBa$_{2}$Mn$_{3}$F$_{11}$.

\section{Conclusion}
%16th paragraph
In conclusion the MDD-type 120$^{\circ}$ structure with an IC modulation was identified 
in NaBa$_{2}$Mn$_{3}$F$_{11}$ by the combination of the neutron diffraction
measurement and magnetic structure analysis.
Classical calculations showed that the MDD interaction is the main perturbative term
for the selection of the magnetic ground state.
To elucidate the precise IC structure and to identify its origin, further investigations, 
for instance single crystal neutron diffraction, are required.
Theoretical calculation including long-range MDD interactions may elucidate the IC structure,
as was the case with the field-induced IC structure in the gadolinium gallium garnet~\cite{PRL97_GGG,PRL114_GGG}. 
Consideration on the interlayer interaction would be also important. 
In addition, the study on magnetic dynamics would be beneficial for the search of exotic 
states induced by the MDD interaction.

\section*{Acknowledgements}
%Acknowledgement
We are grateful to G. J. Nilsen and R. Okuma for helpful discussion.
Travel expenses for the neutron diffraction experiments performed using ECHIDNA at
ANSTO, Australia, and WISH at ISIS, United Kingdom, were supported by the General User
Program for Neutron Scattering Experiments, Institute for Solid State Physics,
The University of Tokyo Proposal (No.~13559 and No.~00499), at JRR-3, Japan Atomic Energy Agency, Tokai, Japan.
S. H. was supported by the Japan Society for the Promotion of  Science through the 
Leading Graduate Schools (MERIT).
This work was partially supported by KAKENHI (Grand No. 15K17701).
T. O. was supported by Ministry of Education, Culture, Sports, Science and Technology
(MEXT) of Japan as a social and scientific priority issue (Creation of new functional devices and high-performance materials to support next-generation industries; CDMSI) 
to be tackled by using post-K computer. 

\begin{table*}[htbp]
\centering
\begin{tabular}{ccllllll}
\hline \hline
IRs & & \multicolumn{6}{c}{Basis Vectors [$m_{a}$ $m_{b}$ $m_{c}$]} \\ 
 & & Mn1 & Mn2 & Mn3 & Mn4 & Mn5 & Mn6 \\ \hline \hline
$\Gamma_{1}$ & $\Psi_{1}$ & [2~0~0] & [0~2~0] & [-2~-2~0]
& [2~0~0] & [0~2~0] & [-2~-2~0] \\
 & & & & \\
$\Gamma_{2}$ & $\Psi_{2}$ & [2~0~0] & [0~2~0] & [-2~-2~0]
& [-2~0~0] & [0~-2~0] & [2~2~0] \\
 & & & & \\
$\Gamma_{3}$ & $\Psi_{3}$ & [1~2~0] & [-2~-1~0] & [1~-1~0] 
& [1~2~0] & [-2~-1~0] & [1~-1~0]  \\
 & $\Psi_{4}$ & [0~0~2] & [0~0~2] & [0~0~2] 
& [0~0~2] & [0~0~2] & [0~0~2] \\
 & & & & \\
$\Gamma_{4}$ & $\Psi_{5}$ & [1~2~0] & [-2~-1~0] & [1~-1~0] 
& [-1~-2~0] & [2~1~0] & [-1~1~0]  \\
 & $\Psi_{6}$ & [0~0~2] & [0~0~2] & [0~0~2] 
 & [0~0~-2] & [0~0~-2] & [0~0~-2] \\
 & & & & \\
$\Gamma_{5}$ & $\Psi_{7}$ & [0.5~0~0] & [0~-1~0] & [-0.5~-0.5~0] 
 & [0.5~0~0] & [0~-1~0] & [-0.5~-0.5~0]  \\
 & $\Psi_{8}$ & [0.5~1.5~0] & [0~0.5~0] & [-0.5~1~0] 
 & [0.5~1.5~0] & [0~0.5~0] & [-0.5~1~0] \\
 & $\Psi_{9}$ & [0~0~1.5] & [0~0~0] & [0~0~-1.5] 
 & [0~0~1.5] & [0~0~0] & [0~0~-1.5] \\
 & $\Psi_{10}$ & [-$\frac{\sqrt{3}}{2}$~0~0] & [0~0~0] & [-$\frac{\sqrt{3}}{2}$~-$\frac{\sqrt{3}}{2}$~0] 
 & [-$\frac{\sqrt{3}}{2}$~0~0] & [0~0~0] & [-$\frac{\sqrt{3}}{2}$~-$\frac{\sqrt{3}}{2}$~0] \\
 & $\Psi_{11}$ & [$\frac{\sqrt{3}}{2}$~$\frac{\sqrt{3}}{2}$~0] & [$\sqrt{3}$~$\frac{\sqrt{3}}{2}$~0] & [$\frac{\sqrt{3}}{2}$~0~0] 
 & [$\frac{\sqrt{3}}{2}$~$\frac{\sqrt{3}}{2}$~0] & [$\sqrt{3}$~$\frac{\sqrt{3}}{2}$~0] & [$\frac{\sqrt{3}}{2}$~0~0] \\
 & $\Psi_{12}$ & [0~0~$\frac{\sqrt{3}}{2}$] & [0~0~-$\sqrt{3}$] & [0~0~$\frac{\sqrt{3}}{2}$] 
 & [0~0~$\frac{\sqrt{3}}{2}$] & [0~0~-$\sqrt{3}$] & [0~0~$\frac{\sqrt{3}}{2}$] \\ 
 & & & & \\
$\Gamma_{6}$ & $\Psi_{13}$ & [0.5~0~0] & [0~-1~0] & [-0.5~-0.5~0] 
 & [-0.5~0~0] & [0~1~0] & [0.5~0.5~0]  \\
 & $\Psi_{14}$ & [0.5~1.5~0] & [0~0.5~0] & [-0.5~1~0] 
 & [-0.5~-1.5~0] & [0~-0.5~0] & [0.5~-1~0] \\
 & $\Psi_{15}$ & [0~0~1.5] & [0~0~0] & [0~0~-1.5] 
 & [0~0~-1.5] & [0~0~0] & [0~0~1.5] \\
 & $\Psi_{16}$ & [-$\frac{\sqrt{3}}{2}$~0~0] & [0~0~0] & [-$\frac{\sqrt{3}}{2}$~-$\frac{\sqrt{3}}{2}$~0] 
 & [$\frac{\sqrt{3}}{2}$~0~0] & [0~0~0] & [$\frac{\sqrt{3}}{2}$~$\frac{\sqrt{3}}{2}$~0] \\
 & $\Psi_{17}$ & [$\frac{\sqrt{3}}{2}$~$\frac{\sqrt{3}}{2}$~0] & [$\sqrt{3}$~$\frac{\sqrt{3}}{2}$~0] & [$\frac{\sqrt{3}}{2}$~0~0] 
 & [-$\frac{\sqrt{3}}{2}$~-$\frac{\sqrt{3}}{2}$~0] & [-$\sqrt{3}$~-$\frac{\sqrt{3}}{2}$~0] & [-$\frac{\sqrt{3}}{2}$~0~0] \\
 & $\Psi_{18}$ & [0~0~$\frac{\sqrt{3}}{2}$] & [0~0~-$\sqrt{3}$] & [0~0~$\frac{\sqrt{3}}{2}$] 
 & [0~0~-$\frac{\sqrt{3}}{2}$] & [0~0~$\sqrt{3}$] & [0~0~-$\frac{\sqrt{3}}{2}$] \\
\hline \hline  
\end{tabular}
\caption{Basis vectors for the space group $R\overline{3}c$ with $\bm k = (0, 0, 0)$. The atoms of the nonprimitive basis are defined according to Mn1: $(0.4438,0,0.25)$, Mn2: $(0,0.4438,0.25)$, Mn3: $(0.5562,0.5562,0.25)$, Mn4: $(0.5562,0,0.75)$, Mn5: $(0,0.5562,0.75)$, Mn6: $(0.4438,0.4438,0.75)$.}
\label{tb:IRq0}
\end{table*}

\begin{table*}[htbp]
\centering
\begin{tabular}{ccll|cll}
\hline \hline
IRs &  & \multicolumn{5}{c}{Basis Vectors [$m_{a}$ $m_{b}$ $m_{c}$]} \\ \hline\hline
 & $\hspace{15mm}$ & Mn1a & Mn1b  
&  $\hspace{15mm}$ & Mn3a & Mn3b \\ \hline
$\Gamma_{1}$ & $\Psi_{1}^{(1)}$ & [1~0~0] & [0~1~0] 
& $\Psi_{1}^{(3)}$ & [1~0~0] & [0~1~0] \\
                     & $\Psi_{2}^{(1)}$ & [0~1~0] & [1~0~0]
& $\Psi_{2}^{(3)}$ & [0~1~0] & [1~0~0] \\
                     & $\Psi_{3}^{(1)}$ & [0~0~1] & [0~0~-1] 
& $\Psi_{3}^{(3)}$ & [0~0~1] & [0~0~-1] \\
\hline
$\Gamma_{2}$ & $\Psi_{4}^{(1)}$ & [1~0~0] & [0~-1~0]  
& $\Psi_{4}^{(3)}$ & [1~0~0] & [0~-1~0] \\
                     & $\Psi_{5}^{(1)}$ & [0~1~0] & [-1~0~0] 
& $\Psi_{5}^{(3)}$ & [0~1~0] & [-1~0~0] \\
                     & $\Psi_{6}^{(1)}$ & [0~0~1] & [0~0~1] 
& $\Psi_{6}^{(3)}$ & [0~0~1] & [0~0~1] \\ \hline
 & & Mn2 & & & Mn4 &  \\ \hline
$\Gamma_{1}$ & $\Psi_{1}^{(2)}$ & [1~1~0] & 
& $\Psi_{1}^{(4)}$ & [1~1~0] & \\
\hline
$\Gamma_{2}$ & $\Psi_{2}^{(2)}$ & [$\frac{1}{\sqrt{3}}~-\frac{1}{\sqrt{3}}~0$]  & 
$\hspace{15mm}$ & $\Psi_{2}^{(4)}$ & [$\frac{1}{\sqrt{3}}~-\frac{1}{\sqrt{3}}~0$] & $\hspace{15mm}$ \\
                     & $\Psi_{3}^{(2)}$ & [0~0~1] & 
& $\Psi_{3}^{(4)}$ & [0~0~1] & \\
\hline \hline  
\end{tabular}
\caption{Basis vectors for the space group $R\overline{3}c$ with $\bm k = (h, h, 0)$. The atoms of the nonprimitive basis are defined according to Mn1a: $(0.4438,0,0.25)$, Mn1b: $(0,0.4438,0.25)$, Mn2: $(0.5562,0.5562,0.25)$, Mn3a: $(0.5562,0,0.75)$, Mn3b: $(0,0.5562,0.75)$, Mn4: $(0.4438,0.4438,0.75)$.}
\label{tb:IRIC}
\end{table*}

\begin{table}[htbp]
\centering
\begin{tabular}{cccccccc}
\hline \hline
$\bm k _{1}$ & $t_{1}$ & $t_{2}$ & $c_{4}^{(1)}$ & $c_{5}^{(1)}$ & $c_{6}^{(1)}$ & $c_{2}^{(2)}$ & $c_{3}^{(2)}$ \\
& -1 & -1 & -2.20(14) &  -2.86(34) & -0.63(34) & 3.33(34) & -1.34(43) \\ \hline \hline
$\bm k _{2}$ & $t_{1}$ & $t_{2}$ & $c_{4}^{(1)}$ & $c_{5}^{(1)}$ & $c_{6}^{(1)}$ & $c_{2}^{(2)}$ & $c_{3}^{(2)}$ \\
& -1 & 1 &  0.82(14) &  1.76(15) & 0.59(24) & -0.23(62) & 2.42(20) \\  
\hline \hline  
\end{tabular}
\caption{Refined coefficients of the basis vectors for the magnetic models with 
$\bm k _{1}$ and $\bm k _{2}$.}
\label{tb:refineIC}
\end{table}

\end{document}